\documentclass[genTeX]{nrc1}

\usepackage{psfrag}
\usepackage{nicefrac}
\usepackage{dcolumn}
\usepackage{graphicx}
\usepackage{epsfig}
\usepackage{amsfonts}
\usepackage{amsmath}
\usepackage{amssymb}
\usepackage{bbold}
\usepackage{bm}
\usepackage{cite}

\volyear{xx}{2006}
\journal{Can. J. Phys.}
\received{xxx}
\accepted{xxx}

\begin{document}

\newcolumntype{d}[1]{D{.}{.}{#1}}
\newcolumntype{.}{D{.}{.}{-1}}

\title{Two--Loop Effects and Current Status of the 
$\bf ^4$He$^{\boldsymbol +}$ Lamb Shift}
\author[U.D.Jentschura]{U. D. Jentschura}
\address{Max--Planck--Institut f\"{u}r Kernphysik, Saupfercheckweg 1,
69117 Heidelberg, Germany. \email{Ulrich.Jentschura@mpi-hd.mpg.de}}
\author[M.Haas]{M. Haas}

\shortauthor{Jentschura and Haas}

\maketitle
\begin{abstract}
We report on recent progress in the treatment of 
two-loop binding corrections to the Lamb shift,
with a special emphasis on $S$ and $P$ states.
We use these and other results in order
to infer an updated theoretical value of the 
Lamb shift in $^4$He$^+$.
\end{abstract}
\begin{resume}
Nous examinons le progr\`{e}s r\'{e}cent concernant 
le traitement des corrections \`{a} l'ordre \'{e}lev\'{e}
des diagrammes \`{a} deux boucles contribuants aux
d\'{e}placement de Lamb, sp\'{e}cialement en ce qui concerne
les \'{e}tats $S$ et $P$. Par cons\'{e}quent, 
on d\'{e}duit de nouvelles 
pr\'{e}cises valeurs th\'{e}oriques pour le 
d\'{e}placement de Lamb en $^4$He$^+$.
\end{resume}

%
%
\section{Introduction}
\label{intro}

\begin{figure}
\begin{center}
\begin{minipage}{14cm}
\begin{center}
\includegraphics[width=1.0\linewidth]{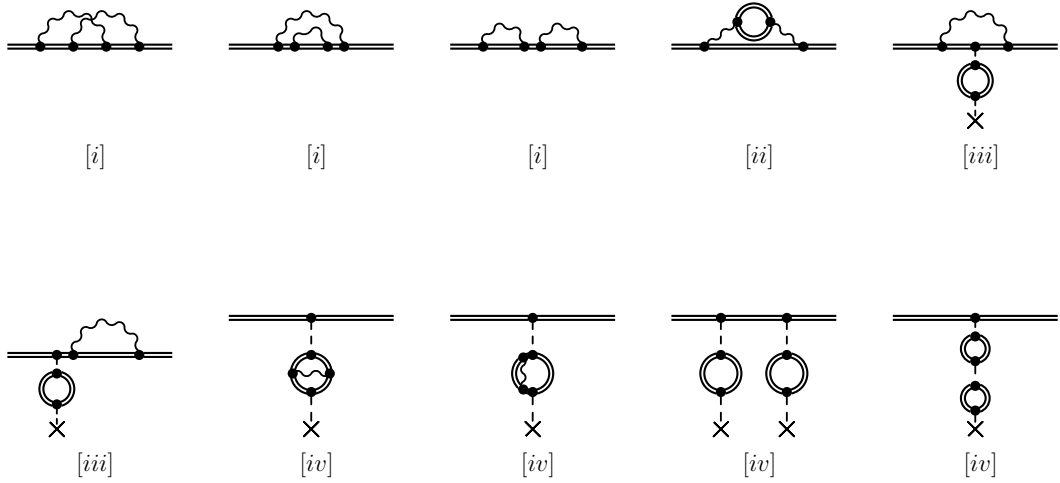}
\end{center}
\caption{\label{fig1} Feynman diagrams
for the two-loop self-energy corrections,
separated into subsets $i$--$iv$ according to 
Ref.~\cite{JeCzPa2005}. Subset $i$ is the pure two-loop self-energy,
subset $ii$ comprises the vacuum-polarization insertion into 
the virtual-photon line of the one-photon self-energy, 
subset $iii$ contains vacuum-polarization corrections 
to the electron line in the one-photon self-energy, and 
subset $iv$ contains remaining vacuum-polarization effects.}
\end{minipage}
\end{center}
\end{figure}

Recently, the higher-order two-loop corrections to the 
Lamb shift have been studied rather intensively, 
both within the $Z\alpha$-expansion
(see~\cite{Pa1993pra,Pa1994prl,EiSh1995,Ka1996,EiGrSh1997,Pa2001,EiGrSh2001,JeCzPa2005} 
and references therein) as well
as within the nonperturbative (in $Z\alpha$) numerical 
approach, as described in 
Refs.~\cite{MaSa1998a,MaSa1998b,YeSh2001,YeInSh2003,%
YeInSh2005jetp,YeInSh2005,YeInSh2006}.
In the current note, we review some recent 
progress for the so-called $B_{60}$ coefficient,
which is generated by the entire gauge-invariant set
of two-loop diagrams, as depicted in Fig.~\ref{fig1}.
We also review some very recent progress~\cite{Je2006} 
regarding the $B_{60}$ coefficient for general 
excited hydrogenic states with nonvanishing angular
momentum, with a special emphasis on states with $P$ symmetry.

Applications of the recent progress to 
high-precision spectroscopy are numerous.
As one example of current interest,
the status of the $^4$He$^+$ Lamb shift
($1S$ and $2S$ states) is summarized, based on the 
recent analytic and numerical results, and 
on information about further known contributions 
to the Lamb shift from the literature
(see in particular~\cite{EiGrSh2001,vWHoDr2001,JeDr2004} 
and references therein).

%
%
\section{Two--Loop Results}
\label{twoloop}

The two-loop energy shift of an atomic level 
in a hydrogenlike atomic system reads
(in units with $\hbar = c = \epsilon_0 = 1$)
\begin{eqnarray}
\label{DefESE2L}
\Delta E^{\rm (2L)}_{\mathrm{SE}} &=& \left(\frac{\alpha}{\pi}\right)^2 \,
\frac{(Z\alpha)^4 m_{\rm e}}{n^3} \, H(Z\alpha)\,,
\end{eqnarray}
where $m_{\rm e}$ is the electron mass,
and $H$ is a dimensionless function.
In the current Section of this 
article, we are primarily concerned with recently 
obtained~\cite{JeCzPa2005,Je2006}
results for the normalized 
(or ``weighted'') difference $H(nS, Z\alpha) - H(1S, Z\alpha)$ of 
$S$ states, whose importance for the 
determination of fundamental constants has been 
stressed in Refs.~\cite{Ka1994,Ka1997,Ka2005}, and for individual $P$ states.

For these states and/or combinations of states,
the first nonvanishing terms in the semi-analytic expansion
of the dimensionless function
$H(Z\alpha)$ in powers for $Z\alpha$ and $\ln(Z\alpha)$
read as follows,
\begin{equation}
\label{defH}
H(Z\alpha) = B_{40}
+ (Z\alpha)^2 \biggl\{ B_{62} \ln^2[(Z\alpha)^{-2}]
+ B_{61} \ln[(Z\alpha)^{-2}] + B_{60} \biggr\}.
\end{equation}
The first index of the $B$ coefficients marks the 
power of $Z\alpha$, whereas the second corresponds to the 
power of the logarithm $\ln[(Z\alpha)^{-2}]$.
For individual $S$ states, we only mention here 
the existence of a $B_{50}$ 
coefficient~\cite{Pa1993pra,EiGrSh1997,Pa1994prl,EiSh1995},
which goes beyond the coefficients listed in (\ref{defH}).

For the normalized difference of $S$ states, we 
have~\cite{Pa2001}
\begin{equation}
\label{B62}
B_{62}(nS) - B_{62}(1S)
= \frac{16}{9} \left( \frac34 + \frac{1}{4 n^2} -
\frac1n + \gamma - \ln(n) \! + \! \Psi(n) \right)\,,
\end{equation}
where $\gamma = 0.577216\dots$ is Euler's constant,
and $\Psi$ is the logarithmic derivative of the Gamma function.
The normalized difference for $B_{61}$ reads~\cite{Pa2001}
\begin{eqnarray}
\label{B61}
B_{61}(nS) - B_{61}(1S)
&=& \frac43\, \left[ N(nS) - N(1S) \right] 
\nonumber\\[2ex]
& & + \left[ \frac{304}{135} - \frac{32}{9} \, \ln(2) \right] \,
\left( \frac34 - \frac{1}{n} + \frac{1}{4\,n^2} 
+ \gamma - \ln(n) \! + \! \Psi(n) \right)\,.
\end{eqnarray}
The normalized difference of the nonlogarithmic term
can be expressed as~\cite{JeCzPa2005}
\begin{equation}
\label{defAnS}
B_{60}(nS) - B_{60}(1S) = b_L(nS) - b_L(1S) + A(n),
\end{equation}
where $A(n)$ is an additional contribution
beyond the $n$-dependence of the two-loop Bethe logarithm
$b_L$. The result for $A$ is~\cite{CzJePa2005,JeCzPa2005},
\begin{align}
\label{An}
& A(n) =
\left( \frac{38}{45} - \frac43\ln(2) \right)  [N(nS) - N(1S)]
- \frac{337043}{129600} - \frac{94261}{21600n} + \frac{902609}{129600n^2}
\nonumber\\
& + \left( \frac{4}{3}
- \frac{16}{9n}
+ \frac{4}{9n^2} \right)  \ln^2 (2)
+ \left( -\frac{76}{45}
+ \frac{304}{135n}
- \frac{76}{135n^2} \right) \ln(2)
+ \left( - \frac{53}{15}
+ \frac{35}{2n}
- \frac{419}{30n^2} \right) 
\nonumber\\
& 
\times \zeta(2)\ln(2)
+ \left( \frac{28003}{10800}
- \frac{11}{2n}
+ \frac{31397}{10800n^2} \right)  \zeta(2)
+ \left( \frac{53}{60}
- \frac{35}{8 n}
+ \frac{419}{120 n^2} \right) \zeta(3)
\nonumber\\
& + \left( \frac{37793}{10800}
+ \frac{16}{9}\ln^2(2)
- \frac{304}{135}\ln (2)
+ 8\zeta(2)\ln(2)
- \frac{13}{3}\zeta(2)
- 2\zeta(3)\right)
\left[ \gamma + \Psi(n) - \ln(n) \right] .
\end{align}
Here, $N(n)$ is a nonlogarithmic term generated by a Dirac-$\delta$
correction to a one-loop Bethe logarithm, as calculated 
in Ref.~\cite{Je2003jpa}. Of course, $\zeta(s) = \sum_{n=1}^\infty n^{-s}$
is the Riemann zeta function.

For $P$ states, we have the known 
results~\cite{Ka1996,CzJePa2005,JeCzPa2005}
\begin{align}
B_{62}(nP) =& \; \frac{4}{27} \frac{n^2 - 1}{n^2} \,,
\nonumber\\[2ex]
B_{61}(nP_{1/2}) =& \; \frac43\, N(nP) + \frac{n^2 - 1}{n^2}
\left(\frac{166}{405} -\frac{8}{27} \, \ln 2 \right) \,,
\nonumber\\[2ex]
B_{61}(nP_{3/2}) =& \; \frac43\, N(nP) +
\frac{n^2 - 1}{n^2}
\left(\frac{31}{405} -\frac{8}{27} \, \ln 2 \right) \, .
\end{align}
The results for the nonlogarithmic terms of $P$ can be inferred
on the basis of Eq.~(8.1) of Ref.~\cite{JeCzPa2005} and the 
two-loop Bethe logarithms for $P$ states (see~\cite{Je2006}
and Table~\ref{table2}),
\begin{subequations}
\label{b60P}
\begin{align}
\label{b60P12}
& B_{60}(nP_{1/2}) = \;
b_L(nP) + \beta_4(nP_{1/2}) + \beta_5(nP_{1/2}) +
\left[ \frac{38}{45} - \frac43\ln(2) \right]  N(nP)
-\frac{27517}{25920} - \frac{209}{288n} 
\nonumber\\[2ex]
& 
+ \frac{1223}{960  n^2}
+ \frac{4}{27}  \frac{n^2 - 1}{n^2}  \ln^2(2)
- \frac{38}{81} \frac{n^2 - 1}{n^2}  \ln(2)
+ \left( \frac{25}{6} + \frac{3}{2  n} - \frac{9}{2 n^2} \right) 
\zeta(2) \ln(2)
\nonumber\\[2ex]
& + \left( -\frac{9151}{10800} -\frac{1}{4n} + \frac{1009}{1200  n^2}
\right)  \zeta(2)
+ \left( -\frac{25}{24} -\frac{3}{8n} + \frac{9}{8  n^2}
\right)  \zeta(3)\,,
\\[2ex]
\label{b60P32}
& B_{60}(nP_{3/2}) =
b_L(nP) + \beta_4(nP_{3/2}) + \beta_5(nP_{3/2}) +
\left[ \frac{38}{45} - \frac43\ln(2) \right]  N(nP) 
-\frac{73321}{103680} + \frac{185}{1152n} 
\nonumber\\[2ex]
&
+ \frac{8111}{25920n^2}
+ \frac{4}{27}  \frac{n^2 - 1}{n^2}  \ln^2(2)
- \frac{11}{81} \frac{n^2 - 1}{n^2}  \ln(2)
+ \left( \frac{299}{80} - \frac{3}{8  n} - \frac{53}{20 n^2} \right) 
\zeta(2) \ln(2)
\nonumber\\[2ex]
& \quad
+ \left( -\frac{24377}{21600} + \frac{1}{16n} - \frac{3187}{3600  n^2}
\right)  \zeta(2)
+ \left( -\frac{299}{320} + \frac{3}{32n} + \frac{53}{80  n^2}
\right)  \zeta(3) \,.
\end{align}
\end{subequations}
In these formulas, $\beta_4$ and $\beta_5$ are low-energy spin-dependent
contributions, defined in Eq.~(4.21) of Ref.~\cite{JeCzPa2005}, 
whose numerical values may be inferred from
one-loop calculations~\cite{Je2003jpa,JeEtAl2003}.

The evaluation of the two-loop
Bethe logarithm for $1S$ and $2S$ has been discussed in
Ref.~\cite{PaJe2003}, and for $3S$--$6S$ in Ref.~\cite{Je2004b60}.
For $1S$ and $2S$, there is no ambiguity in the definition
of the Bethe logarithm, which can roughly be explained
as follows: essentially, the two-loop Bethe logarithm 
results from a renormalized integration over two
photon energies. Both of these integrations are free of singularities
for $1S$ and $2S$.  However, for all higher excited $S$ states and
all $P$ states, one incurs
real (rather than imaginary) contributions to the energy shift
from the product of imaginary contributions due to singularities
along both photon integrations (these are
``squared decay rates'' in the sense of Ref.~\cite{JeEvKePa2002}).
It is thus necessary to make a clear distinction between the
singularity-free, principal-value part $\overline{b}_L$ and
a real part $\delta^2 B_{60}$, which is incurred by
``squared'' (or, more precisely, products of)
imaginary contributions from the pole terms. We write
\begin{equation}
\label{bLdef}
b_L = \overline{b}_L + \delta^2 B_{60}\,,
\end{equation}
where $\overline{b}_L$ is obtained as the nonlogarithmic
energy shift stemming from the nonrelativistic
self-energy, with all integrations carried out by principal
value, and $\delta^2 B_{60}$ is the corresponding
contribution defined in Refs.~\cite{JeEvKePa2002,Je2004b60},
due to squared imaginary parts.
For $3S$--$6S$ states, the above
separation is not really essential, because $\delta^2 B_{60}$
is a numerically marginal contribution as compared to
$\overline{b}_L$ (see Ref.~\cite{Je2004b60}),
and thus $b_L(nS) \approx \overline{b}_L(nS)$ to a
very good approximation.
For $P$ states under investigation here, the distinction (\ref{bLdef}),
surprisingly, is already important (see Table~\ref{table1}).
Final numerical values of the weighted difference 
of $B_{60}$ for $S$ states, and for individual $P$ states,
are summarized in Table~\ref{table2}.

\begin{center}
\begin{table}
\begin{center}
\begin{minipage}{12.0cm}
\begin{center}
\caption{\label{table1}
Total numerical values of the two-loop Bethe logarithms $b_L$
for $S$ and $P$ states, broken down for the 
principal-value contribution $\overline{b}_L$ and the 
squared-decay term $ \delta^2 B_{60}$.}
\begin{tabular}{c@{\hspace{0.3cm}}c@{\hspace{0.7cm}}%
c@{\hspace{0.3cm}}c@{\hspace{0.7cm}}c@{\hspace{0.3cm}}%
c@{\hspace{0.7cm}}c@{\hspace{0.3cm}}c}
\hline
\hline
\rule[-2mm]{0mm}{6mm}
level & $\overline{b}_L$ & $\delta^2 B_{60}$ & $b_L$ & 
level & $\overline{b}_L$ & $\delta^2 B_{60}$ & $b_L$ \\
\hline
$1S$ & $-81.4(3)$ & $0.0$    & $-81.4(3)$ & $-$  & $-$       & $-$      & $-$ \\
$2S$ & $-66.6(3)$ & $0.0$    & $-66.6(3)$ & $2P$ & $-2.2(3)$ & $-0.008$ & $-2.2(3)$ \\
$3S$ & $-63.5(6)$ & $-0.071$ & $-63.6(6)$ & $3P$ & $-2.5(3)$ & $-0.177$ & $-2.7(3)$ \\
$4S$ & $-61.8(8)$ & $-0.109$ & $-61.9(8)$ & $4P$ & $-2.8(3)$ & $-0.243$ & $-3.0(3)$ \\
$5S$ & $-60.6(8)$ & $-0.129$ & $-60.7(8)$ & $5P$ & $-2.8(3)$ & $-0.276$ & $-3.1(3)$ \\
$6S$ & $-59.8(8)$ & $-0.141$ & $-59.9(8)$ & $6P$ & $-2.9(3)$ & $-0.295$ & $-3.2(3)$ \\
\hline
\hline
\end{tabular}
\end{center}
\end{minipage}
\end{center}
\end{table}
\end{center}

\begin{center}
\begin{table}
\begin{center}
\begin{minipage}{12.0cm}
\begin{center}
\caption{\label{table2}
Numerical values for the weighted difference of $B_{60}$
for $S$ states, and for individual $P$ states.}
\begin{tabular}{c@{\hspace{0.3cm}}c@{\hspace{0.7cm}}%
c@{\hspace{0.3cm}}c@{\hspace{0.7cm}}c@{\hspace{0.3cm}}%
c@{\hspace{0.7cm}}c@{\hspace{0.3cm}}c}
\hline
\hline
\rule[-2mm]{0mm}{6mm}
level & $B_{60}(nS)-B_{60}(1S)$ & level & $B_{60}(2P_{1/2})$ & level & $B_{60}(2P_{3/2})$ \\
\hline
$2S$ & $15.1(4)$  & $2P_{1/2}$ & $-1.6(3)$ & $2P_{3/2}$ & $-1.8(3)$ \\
$3S$ & $18.3(7)$  & $3P_{1/2}$ & $-2.0(3)$ & $3P_{3/2}$ & $-2.2(3)$ \\
$4S$ & $20.0(10)$ & $4P_{1/2}$ & $-2.4(3)$ & $4P_{3/2}$ & $-2.5(3)$ \\
$5S$ & $21.2(11)$ & $5P_{1/2}$ & $-2.4(3)$ & $5P_{3/2}$ & $-2.5(3)$ \\
$6S$ & $22.0(11)$ & $6P_{1/2}$ & $-2.5(3)$ & $6P_{3/2}$ & $-2.6(3)$ \\
\hline
\hline
\end{tabular}
\end{center}
\end{minipage}
\end{center}
\end{table}
\end{center}


\begin{center}
\begin{table}[!htb]
\begin{center}
\begin{minipage}{1.0\textwidth}
\caption{\label{ContribsForS}
Contributions to the Lamb shifts of the $1S_{1/2}$ and $2S_{1/2}$ states
in $^4$He$^+$. All equation numbers are connected to the 
contributions as listed in Ref.~\cite{EiGrSh2001}, unless indicated 
otherwise. SE$=$ self-energy, VP$=$ vacuum polarization, 
num.~int.$=$ numerical integration, $m=$ mass of orbiting particle,
for the electron $m=m_{\rm e}$, and $M=$ nuclear mass.}
%
\begin{tabular}{l@{\hspace{0.1cm}}l@{\hspace{0.1cm}}d{5.3}d{5.3}}
\hline
\hline
\multicolumn{1}{l}{\rule[-3mm]{0mm}{8mm}
Order of contribution [$m_{\rm e} c^2$]} &
Equation in Ref.~\cite{EiGrSh2001} &
\multicolumn{1}{c}{${\cal L}(1S)$ \; [MHz] \hspace{0.4cm} $ $} &
\multicolumn{1}{c}{${\cal L}(2S)$ \; [MHz]} \\
\hline
\rule[-1.5mm]{0mm}{5mm}
$ \alpha (Z\alpha)^4 \ln[(Z\alpha)^{-2}] $ & Eq.~(59) [part] &  146\,724.762 &   18\,340.595\\ 
\rule[-1.5mm]{0mm}{5mm}
$ \alpha (Z\alpha)^4 $                     & Eq.~(59) [part] &  -40\,796.296 &   -4\,725.621\\ 
\rule[-1.5mm]{0mm}{5mm}
$ \alpha^2 (Z\alpha)^4 $                   & Eq.~(59) [part] &        16.295 &         2.037\\ 
\rule[-1.5mm]{0mm}{5mm}
$ \alpha^3 (Z\alpha)^4 $                   & Eq.~(59) [part] &         0.029 &         0.004\\ 
\rule[-1.5mm]{0mm}{5mm}
$ \alpha (Z\alpha)^4 $ (muonic vac. pol.)  & Eq.~(63)        &        -0.081 &        -0.010\\ 
\rule[-1.5mm]{0mm}{5mm}
$ \alpha (Z\alpha)^4 $ (hadronic vac. pol.)& Eq.~(65)        &        -0.051 &        -0.006\\ 
\rule[-1.5mm]{0mm}{5mm}
$ \alpha (Z\alpha)^5$ (SE$+$VP)            & Eq.~(73)        &    1\,827.214 &       228.402\\ 
\rule[-1.5mm]{0mm}{5mm}
$\alpha^2(Z\alpha)^5$ (two one-loops) &Eqs.~(74$+$76$+$79$+$80)&       0.492 &         0.061\\ 
\rule[-1.5mm]{0mm}{5mm}
$ \alpha^2 (Z\alpha)^5 $ (two-loop VP)     & Eq.~(75)        &         0.704 &         0.088\\ 
\rule[-1.5mm]{0mm}{5mm}
$ \alpha^2 (Z\alpha)^5 $ (two-loop SE)     & Eq.~(81)        &       -10.709(1) &     -1.339\\ 
\rule[-1.5mm]{0mm}{5mm}
$ \alpha^2 (Z\alpha)^5 $ (two-loop sum)    & sum of 3 above  &        -9.513(1) &     -1.189\\ 
\rule[-1.5mm]{0mm}{5mm}
$ \alpha(Z\alpha)^6 \ln^2[(Z\alpha)^{-2}]$ & Eq.~(84) [part] &       -198.172 &       -24.771\\ 
\rule[-1.5mm]{0mm}{5mm}
$ \alpha (Z\alpha)^6 \ln[(Z\alpha)^{-2}] $ & Eq.~(84) [part] &        123.906 &        16.985\\ 
\rule[-1.5mm]{0mm}{5mm}
$\alpha (Z\alpha)^6 G_{\rm SE}(Z\alpha)$& Ref.~\cite{JeMoSo2001pra} & -82.542 &       -10.621\\ 
\rule[-1.5mm]{0mm}{5mm}
$ \alpha (Z\alpha)^6 G_{\rm VP}(Z\alpha) $ & Ref.~\cite{Mo1996} &      -1.685 &        -0.276\\ 
\rule[-1.5mm]{0mm}{5mm}
$ \alpha (Z\alpha)^6 G_{\rm WK}(Z\alpha) $ & Eq.~(101)       &          0.157 &         0.020\\ 
\rule[-1.5mm]{0mm}{5mm}
$\alpha^2(Z\alpha)^6\ln^3[(Z\alpha)^{-2}]$ & Ref.~\cite{Ka1996} &      -1.153 &        -0.144\\ 
\rule[-1.5mm]{0mm}{5mm}
$\alpha^2(Z\alpha)^6\ln^2[(Z\alpha)^{-2}]$ & Ref.~\cite{Pa2001} &      -0.213 &         0.037\\ 
\rule[-1.5mm]{0mm}{5mm}
$\alpha^2(Z\alpha)^6\ln[(Z\alpha)^{-2}]$   & Ref.~\cite{JeCzPa2005} &   2.666 &         0.279\\ 
\rule[-1.5mm]{0mm}{5mm}
$\alpha^2(Z\alpha)^6$ & Refs.~\cite{JeCzPa2005,YeInSh2005} &           -0.607(211) &   -0.064(26) \\ 
\rule[-1.5mm]{0mm}{5mm}
$ (Z\alpha)^5 m/M $                        & Eq.~(136)      &          17.786 &         2.547\\ 
\rule[-1.5mm]{0mm}{5mm}
$ (Z\alpha)^6 m/M $                        & Eq.~(144)      &          -0.119 &        -0.015\\ 
\rule[-1.5mm]{0mm}{5mm}
$ (Z\alpha)^7 \ln^2(Z\alpha) m/M $         & Eq.~(147)      &          -0.010 &        -0.001\\ 
\rule[-1.5mm]{0mm}{5mm}
$\alpha (Z\alpha)^5 m/M$ & Eq.~(151)$+$Eq.~(46) of Ref.~\cite{EiGrSh2001pra} & -0.112 &   -0.014\\ 
\rule[-1.5mm]{0mm}{5mm}
$\alpha (Z\alpha)^6 \ln^2[(Z\alpha)^{-2}] m/M $ & Eq.~(155) &           0.018 &         0.002\\ 
\rule[-1.5mm]{0mm}{5mm}
$ (Z\alpha)^4 (m/M)^2  $      & Eq.~(15) of Ref.~\cite{PaKa1995} &     -0.053 &        -0.007\\ 
\rule[-1.5mm]{0mm}{5mm}
$ Z\,(Z\alpha)^5 (m/M)^2  $                & approx., Eq.~(152) &       0.019(19) &     0.002(2)\\ 
\rule[-1.5mm]{0mm}{5mm}
Nucl.~size [rel., 1.680(5)\,fm]   & Eq.~(\ref{ResFS}) of this work &  70.865(422)&     8.860(53)\\ 
\rule[-1.5mm]{0mm}{5mm}
Nucl.~size [rel., 1.673(1)\,fm]   & Eq.~(\ref{ResFS}) of this work &  70.275(84) &     8.786(11)\\ 
\hline 
\rule[-1.5mm]{0mm}{5mm}
Sum [1.680(5)\,fm]                         &   &                   107693.112(472)& 13837.031(59)\\
\rule[-1.5mm]{0mm}{5mm}
Sum [1.673(1)\,fm]                         &   &                   107692.522(228)& 13836.957(29)\\
\hline
\hline
\end{tabular}
\end{minipage}
\end{center}
\end{table}
\end{center}


\begin{center}
\begin{table}[!htb]
\begin{center}
\begin{minipage}{1.0\textwidth}
\caption{\label{ContribsForP}
Contributions to the Lamb shifts of the $2P_{1/2}$ and $2P_{3/2}$ states
in $^4$He$^+$. 
As in Table~\ref{ContribsForS}, all equation numbers are connected to the
contributions as listed in Ref.~\cite{EiGrSh2001}, unless indicated
otherwise. The acronyms used for the corrections are also the same
as in Table~\ref{ContribsForS}.}
%
\begin{tabular}{l@{\hspace{0.1cm}}l@{\hspace{0.1cm}}d{5.3}d{5.3}}
\hline
\hline
\multicolumn{1}{l}{\rule[-3mm]{0mm}{8mm}
Order of contribution [$m_{\rm e} c^2$]} &
Equation in Ref.~\cite{EiGrSh2001} &
\multicolumn{1}{c}{${\cal L}(2P_{1/2})$ \; [MHz] \hspace{0.4cm} $ $} &
\multicolumn{1}{c}{${\cal L}(2P_{3/2})$ \; [MHz]} \\
\hline
\rule[-1.5mm]{0mm}{5mm}
$ \alpha (Z\alpha)^4 \ln[(Z\alpha)^{-2}] $ & Eq.~(60) [part] &         0.000 &         0.000\\
\rule[-1.5mm]{0mm}{5mm}
$ \alpha (Z\alpha)^4 $                     & Eq.~(60) [part] &      -206.095 &       200.725\\ 
\rule[-1.5mm]{0mm}{5mm}
$ \alpha^2 (Z\alpha)^4 $                   & Eq.~(60) [part] &         0.414 &        -0.207\\ 
\rule[-1.5mm]{0mm}{5mm}
$ \alpha^3 (Z\alpha)^4 $                   & Eq.~(60) [part] &        -0.003 &         0.002\\ 
\rule[-1.5mm]{0mm}{5mm}
$ \alpha (Z\alpha)^4 $ (muonic vac. pol.)  & Ref.~\cite{vWHoDr2001} & 0.000 &         0.000\\
\rule[-1.5mm]{0mm}{5mm}
$ \alpha (Z\alpha)^4 $ (hadronic vac. pol.)& Ref.~\cite{vWHoDr2001} & 0.000 &         0.000\\ 
\rule[-1.5mm]{0mm}{5mm}
$ \alpha (Z\alpha)^5$ (SE$+$VP)            & Eq.~(73)        &         0.000 &         0.000\\ 
\rule[-1.5mm]{0mm}{5mm}
$\alpha^2(Z\alpha)^5$ (two one-loops) &Eqs.~(74$+$76$+$79$+$80)&       0.000 &         0.000\\
\rule[-1.5mm]{0mm}{5mm}
$ \alpha^2 (Z\alpha)^5 $ (two-loop VP)     & Eq.~(75)        &         0.000 &         0.000\\
\rule[-1.5mm]{0mm}{5mm}
$ \alpha^2 (Z\alpha)^5 $ (two-loop SE)     & Eq.~(81)        &         0.000 &         0.000\\
\rule[-1.5mm]{0mm}{5mm}
$ \alpha^2 (Z\alpha)^5 $ (two-loop sum)    & sum of 3 above  &         0.000 &         0.000\\
\rule[-1.5mm]{0mm}{5mm}
$ \alpha(Z\alpha)^6 \ln^2[(Z\alpha)^{-2}]$ & Eq.~(86) [part] &         0.000 &         0.000\\
\rule[-1.5mm]{0mm}{5mm}
$ \alpha (Z\alpha)^6 \ln[(Z\alpha)^{-2}] $ & Eq.~(86)        &         1.677 &         0.944\\
\rule[-1.5mm]{0mm}{5mm}
$\alpha (Z\alpha)^6 G_{\rm SE}(Z\alpha)$ &Ref.~\cite{JeMoSo2001pra}&  -0.329 &        -0.163\\ 
\rule[-1.5mm]{0mm}{5mm}
$ \alpha (Z\alpha)^6 G_{\rm VP}(Z\alpha) $ & Ref.~\cite{Mo1996}&      -0.022 &        -0.005\\ 
\rule[-1.5mm]{0mm}{5mm}
$ \alpha (Z\alpha)^6 G_{\rm WK}(Z\alpha) $ & Eq.~(101)       &         0.000 &         0.000\\ 
\rule[-1.5mm]{0mm}{5mm}
$\alpha^2(Z\alpha)^6\ln^3[(Z\alpha)^{-2}]$ & Ref.~\cite{Ka1996} &      0.000 &         0.000\\ 
\rule[-1.5mm]{0mm}{5mm}
$\alpha^2(Z\alpha)^6\ln^2[(Z\alpha)^{-2}]$ & Ref.~\cite{Ka1996} &      0.006 &       0.006\\ 
\rule[-1.5mm]{0mm}{5mm}
$\alpha^2(Z\alpha)^6\ln[(Z\alpha)^{-2}]$   & Ref.~\cite{JeCzPa2005}&   0.001 &        -0.001\\
\rule[-1.5mm]{0mm}{5mm}
$\alpha^2(Z\alpha)^6$ & Refs.~\cite{JeCzPa2005,Je2006}       &        -0.001 &        -0.001\\ 
\rule[-1.5mm]{0mm}{5mm}
$ (Z\alpha)^5 m/M $                        & Eq.~(136)       &        -0.138 &        -0.138\\ 
\rule[-1.5mm]{0mm}{5mm}
$ (Z\alpha)^6 m/M $                        & Eq.~(145)       &         0.007 &         0.007\\ 
\rule[-1.5mm]{0mm}{5mm}
$ (Z\alpha)^7 \ln^2(Z\alpha) m/M $         & Eq.~(147)       &         0.000 &         0.000\\
\rule[-1.5mm]{0mm}{5mm}
$\alpha (Z\alpha)^5 m/M$ & Eq.~(151)                         &  0.000        & 0.000\\
\rule[-1.5mm]{0mm}{5mm}
$\alpha (Z\alpha)^6 \ln^2[(Z\alpha)^{-2}] m/M $ & Eq.~(155)  &         0.000 &         0.000\\
\rule[-1.5mm]{0mm}{5mm}
$ (Z\alpha)^4 (m/M)^2  $      &   E.g., Ref.~\cite{SaYe1990} &      0.002 &        -0.001\\ 
\rule[-1.5mm]{0mm}{5mm}
$ Z\,(Z\alpha)^5 (m/M)^2  $                & approx., Eq.~(152) &      0.000 &         0.000\\ 
\rule[-1.5mm]{0mm}{5mm}
Nucl.~size                      & Eq.~(\ref{ResFS}) of this work &     0.000 &         0.000\\ 
\hline
\rule[-1.5mm]{0mm}{5mm}
Sum                                        &                 &      -204.481 &       201.168\\
\hline
\hline
\end{tabular}
\end{minipage}
\end{center}
\end{table}
\end{center}

%
%
\section{Status of the $^4$He$^+$ Lamb Shift}
\label{helamb}

In the current section (we keep units with $\hbar = c = \epsilon_0 = 1$),
we would like to use the results described above, in order to infer the current
theoretical status of the Lamb shift of $1S$ and $2S$ in the
${}^4$He$^+$ ion. Before we start our actual discussion, however,
we should remember that an ideal way to carry out a related calculation
would involve a full-featured least-squares adjustment
according to Ref.~\cite{JeEtAl2005}, which includes all available
data from relevant high-precision experiments (see~\cite{vWHoDr2001})
and which, in principle, allows for a deduction of the nuclear
charge radius. In order to infer an approximate theoretical
prediction, though, one has to use a charge radius obtained 
from other sources, and we intend to follow this different 
route in the current work.

We partly base our evaluation on 
Refs.~\cite{vWHoDr2001,JeDr2004,EiGrSh2001}
and choose a format as in Table~1 of Ref.~\cite{vWHoDr2001}.
In the evaluations described in 
Tables~\ref{ContribsForS} and~\ref{ContribsForP}, the 2002 CODATA
values of the fundamental constants~\cite{MoTa2005} were used.

For the Lamb shift ${\cal L}$,
we use the implicit definition~\cite{SaYe1990,PaKa1995}
\begin{equation}
\label{defElamb}
E = m_{\rm r} \left[ f(n,j)-1 \right] - \frac{m_{\rm r}^2}{2 (m_{\rm e} + M)}
\left[ f(n,j) - 1 \right]^2 + {\cal L} + E_{\rm hfs}\,.
\end{equation}
Here, $E$ is the energy level of the bound two-body system
under investigation, and $f(n,j)$ is the dimensionless Dirac energy.
E.g., we have $f(1,1/2) = f(1S) = \sqrt{1 - (Z\alpha)^2}$,
and $f(2,1/2) = f(2S) = \sqrt{\frac12\,(1 + \sqrt{1 - (Z\alpha)^2})}$
for the $1S$ and $2S$ states, respectively.
Furthermore, $m_{\rm r}$ is the reduced mass of the system,
$M$ is the nuclear mass, and $E_{\rm hfs}$ is the energy
shift due to hyperfine effects, which are absent for the 
spinless ${}^4 {\rm He}$ nucleus.

In order to avoid confusion, we would like to include a few clarifying
words regarding specific entries in 
Tables~\ref{ContribsForS} and~\ref{ContribsForP}.
In general, we have added the factor $\alpha$ 
to all scales for the contributions listed in 
Tables~\ref{ContribsForS} and~\ref{ContribsForP}.
giving all contributions with an overall scaling of 
$m_{\rm e}c^2$, in contrast to $\alpha\, m_{\rm e}c^2$,
which had been used in Ref.~\cite{vWHoDr2001}. 
Regarding the contribution of order $ (Z\alpha)^4 \ln[(Z\alpha)^{-2}] $
in the first row of Table~\ref{ContribsForS},
it is worthwhile to note that this term
represents the leading logarithm of the Lamb shift, given by
\begin{equation}
\label{q}
\alpha \, \frac{(Z\alpha)^4 m_{\rm e}}{n^3} \, 
\ln[(Z\alpha)^{-2}] \, \left( \frac{ m_{\rm r} }{ m_{\rm e} } \right)^3 \,.
\end{equation}
Here, $m_{\rm r}$ is the reduced mass of the system,
given by 
$m_{\rm r} = m_{\rm e} m_{\rm N}/(m_{\rm e} + m_{\rm N})$,
where $m_{\rm N}$ is the mass of the nucleus.
The reduced-mass dependence of the argument of the logarithm
itself, $\ln[(Z\alpha)^{-2}] \to \ln[(Z\alpha)^{-2} \, m_{\rm e}/m_{\rm r} ] 
= \ln[(Z\alpha)^{-2}] +  \ln[m_{\rm e}/m_{\rm r} ]$,
is being included here and in Ref.~\cite{vWHoDr2001}
into the nonlogarithmic term of order
$ \alpha (Z\alpha)^4 $.
One might wonder why there is a theoretical uncertainty associated to this
contribution at all. The reason is that the most accurate
theoretical value for the quantity (\ref{q}) is obtained
by expressing it in terms of the 2002 CODATA Rydberg
constant, which has a relative uncertainty of $6.6 \times 10^{-12}$,
and the 2002 CODATA fine-structure constant,
which has a relative uncertainity of $3.3 \times 10^{-12}$.
The latter is responsible for the small theoretical uncertainty
of the leading logarithmic contribution to the Lamb shift
of the $1S$ level.
The term of order $ \alpha (Z\alpha)^5 $ contains both 
contributions from the self-energy and the vacuum
polarization, as indicated by the explanatory note ``SE$+$VP.''

The term of order $\alpha (Z\alpha)^6 \ln^2[(Z\alpha)^{-2}] $
corresponds to the self-energy coefficient $A_{62}$,
as given e.g.~in Ref.~\cite{Pa1993}, and 
the indicated term of order $\alpha (Z\alpha)^6 \ln[(Z\alpha)^{-2}] $
is the sum of a self-energy and a vacuum-polarization
contribution in this order. Note that the latter
distinction differs from the one used in Ref.~\cite{vWHoDr2001},
where the term of order $\alpha (Z\alpha)^6 \ln^2[(Z\alpha)^{-2}] $
denotes the sum of a double logarithmic, and a single logarithmic 
{\em self-energy} contribution, and the term of order
$\alpha (Z\alpha)^6 \ln[(Z\alpha)^{-2}] $ was reserved
exclusively for the {\em vacuum-polarization} contribution 
in this order. 
The values of the self-energy remainder $G_{\rm SE}(Z\alpha)$ for
$S$ and $P$ states are listed in~\cite{JeMoSo2001pra,JeMo2004pra}. 
The vacuum polarization remainder function $G_{\rm VP}(Z\alpha)$
is taken from~\cite{Mo1996} and corresponds exclusively 
to the Uehling part of the one-loop vacuum polarization.
It might be worthwhile to point out that at
the current level of accuracy, it is entirely sufficient
to consider the vacuum-polarization higher-order 
remainder for $P$ states via the formula
\begin{equation}
\Delta E_{\rm VP}(nP_j) = \frac{\alpha}{\pi} \,
\frac{(Z\alpha)^6 m_{\rm e}}{n^3} \, 
\left( \frac{ m_{\rm r} }{ m_{\rm e} } \right)^3 \,
\left\{A^{\rm VP}_{60}(nP_j) +
(Z\alpha) \, A^{\rm VP}_{70}(nP_j) \right\} \,,
\end{equation}
where the analytic coefficients read
\begin{subequations}
\label{ResVP}
\begin{align}
& A^{\rm VP}_{60}(nP_{1/2}) = - \frac{3}{35}\, \frac{n^2 - 1}{n^2}\,, 
\qquad
A^{\rm VP}_{60}(nP_{3/2}) = - \frac{2}{105}\, \frac{n^2 - 1}{n^2}\,,
\\[2ex]
& A^{\rm VP}_{70}(nP_{1/2}) = \frac{41 \pi}{2304}\, \frac{n^2 - 1}{n^2}\,,
\qquad
A^{\rm VP}_{70}(nP_{3/2}) = \frac{7 \pi}{768}\, \frac{n^2 - 1}{n^2}\,.
\end{align}
\end{subequations}
These have been obtained in Refs.~\cite{JeSoMo1997,EiGrSh2001,Je2006}
for general principal quantum number $n$.

For the $B_{62}$-term for $S$ states of order
$\alpha^2(Z\alpha)^6\ln^2[(Z\alpha)^{-2}]$, the result from
Ref.~\cite{Pa2001} was used, which supersedes the 
estimate given in Eq.~(101) of Ref.~\cite{EiGrSh2001}.
For the analytic $B_{61}$-term of order
$\alpha^2(Z\alpha)\ln[(Z\alpha)^{-2}]$, the result given in
Ref.~\cite{JeCzPa2005} provides the most recent value.

For the $B_{60}$ coefficient corresponding to the 
nonlogarithmic term of order $\alpha^2 (Z\alpha)^6$
for the ground state,
two mutually contradictory results of $-61.6 \pm 15\,\%$ 
(Ref.~\cite{PaJe2003}) and $-127 \pm 30\,\%$ 
(Ref.~\cite{YeInSh2005}) have been reported. The latter is
based on an extrapolation of an all-order (in $Z\alpha$) 
numerical calculation. Note, however, that there is a known missing
piece in the analytic result reported in Ref.~\cite{PaJe2003},
which is currently under study and which will need to be
evaluated before final conclusions can be drawn. 
M. Eides~\cite{EiPriv2006}
therefore suggested that a valid interim way of estimating the uncertainty
of $B_{60}$ would consist in taking the arithmetic mean 
of these two results, and taking the half difference 
as an estimate for the theoretical uncertainty.
This uncertainty would comprise all higher-order analytic terms,
as it involves a comparison to a nonperturbative (in $Z\alpha$) calculation.
For the $2S$ state, we can use the result for $1S$ and add the 
weighted difference listed in Table~\ref{table2}.
For $P$ states, the results reported in Sec.~\ref{twoloop}
of this paper (see also~\cite{Je2006}) provide enough information
to eliminate all theoretical uncertainty at the current level
of accuracy.

Finally, let us remark that 
a term of order $ \alpha^2 (Z\alpha)^7 \ln^2[(Z\alpha)^{-2}]$
could be estimated in principle on the basis of taking a ``local'' Lamb-shift
potential that corresponds to the self-energy part of 
$A_{50}$, namely, 
\begin{equation} 
\delta V = 4 \, \alpha \, (Z\alpha)^2 \left[\frac{139}{128} - 
\frac12\, \ln(2) \right]\, \frac{\pi \delta^3(r)}{m_{\rm e}^2},
\end{equation}
taken as an input for a Dirac-$\delta$ induced correction to the 
one-loop self-energy. The result of this approach could alternatively
be used as an uncertainty estimate for all the higher-order terms.
This procedure leads to the estimate 
\begin{equation} 
B_{72}(nS) = \pm \frac83\, \pi \,\left[\frac{139}{128} -
\frac12\, \ln(2) \right]\,,
\end{equation} 
For $1S$, this leads to a value 
of $\pm 0.337$~MHz for the higher-order two-loop remainder. 
However, the uncertainty due to the $B_{72}$-contribution is already
contained in the uncertainty of the remainder term 
of order $\alpha^2 (Z\alpha)^6$, because in determining
the uncertainty of $B_{60}$, a comparison was made to a 
nonperturbative numerical calculation for higher nuclear charge
numbers. The latter necessarily contains all contributions from the 
$B_{72}$ term and all higher-order remainders. The above result
of $\pm 0.337$~MHz therefore likely overestimates the uncertainty
due to the higher-order two-loop remainder and is mentioned 
here only for illustrative purposes.

Concerning radiative-recoil corrections, we note that the 
discrepancy between~\cite{Pa1995} and~\cite{BhGr1987} concerning radiative
insertions into the electron line in the $ \alpha (Z\alpha)^5 m/M $
radiative recoil correction was resolved in~\cite{EiGrSh2001pra}. The
analytical result from that work was used. According to 
Eq.~(46) of Ref.~\cite{EiGrSh2001pra},
the coefficient multiplying the non-vacuum-polarization part of order 
correction of order $ \alpha (Z\alpha)^5 m/M $ is $-1.32402796\dots/n^3$.

The nuclear spin in $^4$He$^+$ is different as compared
to atomic hydrogen. The former is a spin-$1/2$--spin-zero system, whereas the
latter is a spin-$1/2$--spin-$1/2$ system.
Recoil corrections of first order in the mass ratio are
unaffected by the different spin of the nucleus as compared
to hydrogen. However, recoil terms of order $ Z (Z\alpha)^5 (m/M)^2  $ are 
nuclear spin-dependent.
Without carrying out a detailed analysis, we approximately
calculate the nuclear self-energy effects of
order  $ Z (Z\alpha)^5 (m/M)^2  $ by leaving out the Pauli
form-factor correction from Eq.~(153) of Ref.~\cite{EiGrSh2001},
which is certainly absent for a spinless nucleus,
and we conservatively take the Dirac form factor contribution as an 
uncertainty, while we note that a more detailed analysis would be of interest
and currently lacking in the literature.

Finally, we add the nuclear-spin dependent correction listed in 
Eq.~(15) of Ref.~\cite{PaKa1995},
\begin{equation}
\Delta E = -\frac12 \, 
\frac{(Z\alpha)^4 m_{\rm e}}{n^3} \, 
\left( \frac{m_{\rm e}}{M} \right)^2 \,
\delta_{l0} \,,
\end{equation}
which is of second order in the mass ratio,
for the spin-$1/2$--spin-zero system under investigation.
This term is connected to the absence of the zitterbewegung term
in the Breit Hamiltonian for a spinless nucleus.
For $P$ states, we also add terms of order 
$(Z\alpha)^4 \left( \frac{m_{\rm e}}{M} \right)^2$, which do not
depend on the zitterbewegung term and are given, e.g.,
in Ref.~\cite{SaYe1990}.

Concerning the nuclear-size correction, we would like to 
mention that a full integration of a nuclear potential
with a fully relativistic wave function (e.g., within a hard-sphere 
approximation) turns out to be quite essential to obtain
reliable values for this correction. We have carried out such an integration
in the current investigation with the full
Dirac wave functions and obtain results in agreement with 
Ref.~\cite{Sh1993}. 
Results obtained with 
two different values for the two different root-mean-square charge 
radii of $1.673(1)\,{\rm fm}$~\cite{BoRi1978} and 
$1.680(5)\,{\rm fm}$~\cite{SiPriv2006} are given in Table~\ref{ContribsForS}.
Note that the former charge radius of $1.673(1)\,{\rm fm}$ has been 
questioned (see e.g.~\cite{Co1982pra,BrZa1990}).
The uncertainty due to the shape of the nuclear charge 
distribution can be estimated to be much smaller
than the uncertainty due to the nuclear size, based on experience
with highly charged ions~\cite{BeEtAl1997pla}.
For the $^4$He-nucleus, a spherically symmetric model is well justified (closed
shell, spin $I\!=\!0$). 

The nuclear-size correction $\Delta E_{\rm fs}(nS)$ 
and $\Delta E_{\rm fs}(nP_j)$, 
for low nuclear charge numbers, can be approximated
very well by the first few terms of an expansion in the two
small parameters $Z\alpha$ and $m \langle r^2 \rangle^{1/2}$, with the result
\begin{subequations}
\label{ResFS}
\begin{align}
\label{ResFSa}
\Delta E_{\rm fs}(nS) =&
\frac{(Z\alpha)^4 m^3 R^2}{n^3}
\, \left[ \frac25 -
\frac13 \, (Z\alpha \, m \, R) +
\right. 
\nonumber\\[2ex]
& \left. + (Z\alpha)^2 \left\{ -\frac25\,
\left[ \ln\left( \frac{2 \, Z\,\alpha\,m\, R}{n} \right) + 2 \gamma 
+ \Psi(n) \right] + \frac{227}{150} + \frac{2}{5\,n} - \frac{9}{10 n^2}  
\right\} \right] \,,
\\[2ex]
\label{ResFSb}
\Delta E_{\rm fs}(nP_j) =&
\frac{1}{10} \, \frac{(Z\alpha)^6 m^3 R^2}{n^3} \,
\frac{n^2 - 1}{n^2} \, \delta_{j,1/2}\,.
\end{align}
\end{subequations}
Here, $\gamma = 0.577216\dots$ is Euler's constant, $\Psi(n)$ is the
logarithmic derivative of the Gamma function, and
$R$ is the radius of the nucleus in a hard-sphere model, which is
related to the root-mean-square radius $\langle  r^2 \rangle^{1/2}$ 
by the following formula [see Eq.~(7) of Ref.~\cite{Sh1993}],
\begin{equation}
\label{r0corrected}
R = \sqrt{\frac{5}{3}} \, \langle r^2 \rangle^{1/2}
\, \sqrt{ 1- \frac34\, (Z\alpha)^2 \,
\left\{ \frac{3}{25} \, \frac{\langle r^4 \rangle}{\langle r^2 \rangle^2} -
\frac17 \right\} } \,.
\end{equation}
For the $2P_{1/2}$ state, we obtain an upward finite nuclear-size 
energy shift of $353\,{\rm Hz}$ which is barely significant on the kHz level 
(see Table~\ref{ContribsForP}).

The results in Eqs.~(\ref{ResFS}) and~(\ref{r0corrected}) have been
obtained in the approximation of an infinitely heavy nucleus, and with 
exact Dirac wave functions for a point nucleus. Both of these 
approximations should be valid for $^4{\rm He}^+$. In addition,
it should be noted that both the results
given in Eqs.~(\ref{ResFS}) and~(\ref{r0corrected}) are in excellent
numerical agreement with a full numerical integration of the 
finite-size potential with Dirac wave functions. The linear 
correction term 
in $R$, i.e. the term $-\frac13 \, (Z\alpha \, m \, R)$ in 
Eq.~(\ref{ResFSa}), is a consequence of the exponential factor
$\approx \exp(-Z\alpha m r/n)$ in the wave function, which should not
be ignored in the evaluation of the finite-size effect, although this 
effect is primarily sensitive to the probability density 
at the origin (at the nucleus).
Any further effects that influence the finite-size effect
like nuclear polarization are here absorbed into the 
uncertainty of the nuclear radius (see, e.g., Ref.~\cite{Fr1979}
for an illustrative discussion of some of the further aspects that 
are relevant to the finite-size effect).

%
%
\section{Conclusions}
\label{conclu}

In Sec.~\ref{twoloop}, we summarize recent theoretical results
for the higher-order two-loop binding corrections to the Lamb shift.
These results are used,
in Sec.~\ref{helamb}, to infer updated values for the Lamb shift
of low-lying states of the $^4$He$^+$ ion.
Some of the analytic coefficients used in the evaluation
are given in Eqs.~(\ref{ResVP}) and~(\ref{ResFS}).
The analytic expansion of the nuclear finite-size correction~(\ref{ResFS})
might be useful in other contexts.

The recent progress in the field has allowed for an improvement
of the theory beyond the limits set by the leading-order
effects, and for some of the most accurate predictions in 
all of theoretical physics. In particular, we reemphasize that
for $P$ states, the results reported in Sec.~\ref{twoloop}
of this paper (see also~\cite{Je2006}) provide sufficient information
to eliminate all theoretical uncertainty at the kHz level for the 
Lamb shift in the $^4$He$^+$ ion.

%
%
\section*{Acknowledgments}

Helpful conversations with Krzysztof Pachucki, Vladimir M. Shabaev,
Vladimir A. Yerokhin and Peter J. Mohr are gratefully acknowledged.
U.D.J. acknowledges support from the Deutsche Forschungsgemeinschaft
(Heisenberg program).

\end{document}